\pdfoutput=1
\documentclass[12pt]{article}
\usepackage{graphicx}
\usepackage{amssymb,amsfonts,amsmath}

\def\lsim{\mathrel{\rlap{\lower3pt\hbox{\hskip1pt$\sim$}}
     \raise1pt\hbox{$<$}}} 
\def\gsim{\mathrel{\rlap{\lower3pt\hbox{\hskip1pt$\sim$}}
     \raise1pt\hbox{$>$}}} 
\newcommand{\beq}{\begin{equation}}
\newcommand{\eeq}{\end{equation}}
\newcommand{\bea}{\begin{eqnarray}}
\newcommand{\eea}{\end{eqnarray}}

\begin{document}
\title{Recent thermodynamic results from lattice QCD analyzed within a quasi-particle model}
\author{Salvatore Plumari$^{a,b}$, Wanda M. Alberico$^{c,d}$, Vincenzo Greco$^{a,b}$, Claudia Ratti$^{c,d}$\\
$^a$ \small{\it Department of Physics and Astronomy, University of Catania,}\\ 
\small{\it Via S. Sofia 64, I-95125 Catania (Italy)}\\
$^b$ \small{\it Laboratorio Nazionale del Sud, INFN-LNS, Via S. Sofia 63, I-95125 Catania (Italy)}\\
$^c$ \small{\it Dipartimento di Fisica Teorica, Universit\`a degli Studi di Torino}\\
\small{\it via P. Giuria 1, I-10125 Torino (Italy)}\\
$^d$ \small{\it INFN, Sezione di Torino}}
\maketitle
\begin{abstract}
The thermodynamic behavior of QCD matter at high temperature is currently studied by lattice QCD
theory. The main features are the fast rise of the energy density $\epsilon$ around the critical
temperature $T_c$ and the large trace anomaly of the energy momentum tensor
$\langle \Theta_\mu^\mu \rangle=\epsilon - 3 P$ which 
hints at a strongly interacting system. 
Such features can be accounted for by employing a massive quasi-particle model with a temperature-dependent bag constant.
Recent lattice QCD calculations with physical quark masses by the Wuppertal-Budapest group
show a slower increase of $\epsilon$ and a smaller
$\langle\Theta_\mu^\mu\rangle$ peak with respect to previous results from the hotQCD collaboration. 
We investigate the implications of such differences from the point of
view of a quasi-particle model, also discussing light and strange quark number susceptibilities. 
Furthermore, we predict the impact of these discrepancies on the temperature-dependence of the transport properties of matter, like the shear and bulk viscosities.
\end{abstract}
PACS: 12.38.Mh, 25.75-q, 52.25.Fi

\section{Introduction}
QCD at very high temperature $T$ is expected to be weakly coupled \cite{Collins:1974ky}
and provides perturbative screening of the charge \cite{Shuryak:1977ut},
thus being called Quark-Gluon Plasma (QGP). 
Creating and studying this phase of matter in the laboratory has been the goal of experiments at CERN SPS and 
at the Relativistic Heavy Ion Collider (RHIC) facility in Brookhaven
National Laboratory, recently continued by the ALICE, ATLAS and CMS Collaborations
at the Large Hadron Collider (LHC). These experiments
provide the unique possibility of quantifying 
the properties of the deconfined phase of QCD. 
At the same time, lattice calculations on QCD thermodynamics are now reaching 
unprecedented levels of accuracy, with 
simulations at the physical quark masses and several values of the lattice cutoff,
which allows to keep lattice artifacts under control. The information that can be obtained from these
two sources can shed light on the features of QCD matter under extreme conditions, one of the
major challenges of the physics of strong interactions. 

As soon as the first lattice results on the QCD equation of state became available, several attempts
have been made in order to interpret them in terms of the appropriate effective degrees of freedom.

There have been many efforts to discuss QCD thermodynamics at high temperatures by means
of pertubation theory. For the pressure, the calculations have been extended up to $g^6\log g$ \cite{Kajantie:2002wa}, but
besides strong fluctuations from one order to the other, the final result agrees with lattice data only for $T\geq 5 \, T_c$. Moreover, if one concentrates on the interaction
effects by calculating the interaction measure $\langle\Theta_\mu^\mu\rangle=\epsilon - 3 P$,
the pertubative expansion completely fails to reproduce the lattice data. In order to overcome this problem, resummation schemes have been proposed, based for example on the Hard Thermal Loop (HTL) approach \cite{HTL1,HTL2,HTL3,HTL4,HTL5,HTL6,HTL7,HTL8} or 
on dimensionally reduced screened perturbation theory (DRSPT) \cite{DR1,DR2,DR3}.

More recently within a NNLO in the HTL perturbation theory \cite{Andersen_HTLpt} it has been shown that an agrement with the lattice results  can be obtained
also for the trace anomaly down to  $T \simeq 2\,T_c$. At these high temperatures, the HTL 
approach motivates and justifies a picture of weakly interacting quasi-particles, as 
determined by the HTL propagators.

A particularly intriguing possibility is to bring the description of lattice results in terms of quark
and gluon quasi-particles down to $T_c$. Their effective masses are generated through the interaction among the basic constituents
\cite{Gorenstein:1995vm,LH1998,PC05,PKPS96}: if a large part of the interaction can be included into the
effective masses, quasi-particles move freely or interact only weakly. This approach can be very useful in
phenomenological hadronization models \cite{coal} and also for a description of the QCD medium in the vicinity of the
phase transition, where perturbative methods cannot be used.
The quasi-particle model (QPM) was successfully applied to quantitatively describe lattice QCD results of equilibrium thermodynamics such as the EoS and related quantities \cite{qpm}: by a fit to lattice QCD thermodynamics, the quasi-particle properties such as their effective masses can be obtained.

Recently, new lattice results for the equation of state of QCD with 2+1 dynamical flavors have become 
available \cite{Borsanyi:2010cj}. These simulations have been performed at the physical quark masses and for
several lattice spacings, and a continuum estimate of the bulk thermodynamic quantities has been given. These
new results show several differences with respect to the previously available ones \cite{hotQCD2,hotQCD} (a slower increase of $\epsilon$ and a smaller
$\langle\Theta_\mu^\mu\rangle$ peak),
which stimulates the question whether and how the properties of quark and gluon quasi-particles are affected by these features.

In the present paper we assume that it is possible to describe the QCD medium above the deconfinement phase
transition in terms of quark and gluon quasi-particles with an effective mass. 
Our purpose is to check how the properties of these quasi-particles change, following the differences between
the available lattice data for QCD thermodynamics. After performing a fit
to the available lattice data, we will obtain the quark and gluon temperature-dependent masses, which are
related to each other through the coupling constant, and a temperature-dependent bag constant which is
necessary for thermodynamic consistency. Once the thermodynamic properties of the medium are determined, we
will extract dynamical properties such as bulk and shear viscosities and relaxation time. The model predictions for quark-number susceptibilities are also discussed.

\section{The Quasi-Particle Model}
A successful way to account for non-perturbative dynamics is a quasi-particle approach,
in which the interaction is encoded in the quasi-particle masses. The mass of the particles can be viewed as
arising from the energy contained in a strongly coupled volume determined by the correlation range of
the interaction.
Once the interaction is accounted for in this way, the quasi-particles behave like a free gas of massive constituents.
The model is usually completed by introducing a finite bag pressure that can account for further
non-perturbative effects and could be directly linked to the gluon condensate at least
in the pure gauge case \cite{Castorina:2011ja}.
It is already very well known that, in order  to be able to describe
the main features of lattice QCD thermodynamics,
a temperature-dependent mass has to be considered. This also implies that the bag constant has to be
temperature-dependent, in order to ensure thermodynamic consistency.
In fact, when a temperature-dependent mass is included in the pressure, its derivative with respect to the
temperature will produce an extra term in the energy
density, which does not have the ideal gas form. A solution to this problem has been proposed in
\cite{Gorenstein:1995vm}: 
pressure and energy density contain additional medium contributions, which we will call $B(T)$, and which cure
the problem of thermodynamic consistency, as we will see in the following. In this formulation, the entropy density
preserves the ideal gas form. 

The temperature-dependent effective mass for quarks and gluons 
can be evaluated in a perturbative
approach that suggests the following relations~\cite{LH1998,PC05,PKPS96}:
\begin{eqnarray}
& &  m_g^2 = \frac{1}{6} g^2
  \left[
    \left( N_c+ \frac12\, n_{\!f} \right) T^2
    + \frac{N_c}{2\pi^2} \sum_q \mu_q^2
    \right] \, ,
  \nonumber \\
& &  m_{u,d}^2 = \frac{N_c^2-1}{8N_c} g^2 \, 
  \left[ T^2+\frac{\mu_{u,d}^2}{\pi^2} \right] \, ,
\label{mqg}
\end{eqnarray}
where $n_f$ is the number of flavors considered (2 or 3 in this work, see the following), $N_c$ is the number of colors, $m_{u,d}$ is the
mass of the light quarks and $\mu_q$ is the chemical potential of the generic flavor $q$ considered.

The coupling $g$ is generally temperature-dependent. However, as mentioned in the introduction, the calculation of such a $T$-dependence by means of perturbation theory does not allow to have a good description of lattice QCD thermodynamics. Therefore, usually $g(T)$ is left as a function to be determined through the fit to lattice QCD data. 
This will be also our methodology, but we will consider lattice QCD data for $SU(3)$ in flavor
space, hence for the strange quark mass we will use the following relation: 

\begin{eqnarray}
& &  m_{s}^2-m_{0s}^2 = \frac{N_c^2-1}{8N_c} g^2 \, 
  \left[ T^2+\frac{\mu_s^2}{\pi^2} \right] \, ,
\label{mstrange}
\end{eqnarray}
where $m_s$ is the total mass of the strange quark while $m_{0s}$ is the current quark mass.
We see that Eq. (\ref{mstrange}) reduces to the second equation in (\ref{mqg})
when $m_{0s}=0$.
The effect of a finite current quark mass is generally negligible for the description of
the energy density, pressure, entropy but can play a role in the evaluation of 
the strange
quark number susceptibility, see Section \ref{qsusc}.

The pressure of the system can then be written as the sum of independent contributions coming
from the different constituents, which have a $T$-dependent effective mass, plus a bag constant:
\begin{eqnarray}
  P_{qp}(m_u, m_d, ..., T)& = &
  \sum_{i=u,d,s,g}  d_i \int \frac{d^3p}{(2 \pi)^3} \frac{\vec{p}^2}{3 E_i(p)} f_i(p) - B(T) \  ,
\label{pressure}
\end{eqnarray}
where $f_i(p)=[1 \mp \exp{(\beta E_i(p)}]^{-1}$ are the Bose and Fermi distribution functions, with $E_i(p)=\sqrt{\vec{p}^{\,2}+m_i^2}$; $d_i=2 \times 2 \times N_C$ for quarks and $d_i=2 \times (N_C^2-1)$ for gluons.
 
In order to have thermodynamic consistency, the following relationship has to be satisfied:
\begin{equation}
\label{consistenza}
  \bigg(\frac{\partial P_{qp}}{\partial m_i} \bigg)_{T,\mu}=0 \ , \ \ i=u,d,\dots \ \ ,
\end{equation}
which gives rise to a set of equations of the form
\begin{equation}
  \frac{\partial B}{\partial m_i} + d_i \int \frac{d^3p}{(2 \pi)^3} \frac{m_i}{E_i} f_i(E_i) = 0 \ .
  \label{gap}
\end{equation}
Only one of the above equations is independent, since the masses of the constituents all depend on the
coupling $g$ through relationships of the form: $m_i(T,\mu=0)=\alpha_i g(T)T$, 
where $\alpha_i$
are constants depending on $N_c$ and $N_f$ according to Eqs. (\ref{mqg}).
The energy density of the system is then obtained from the pressure through the thermodynamic
relationship $\epsilon(T)=T dP(T)/dT-P(T)$ and will have the form
\begin{eqnarray}
  \epsilon_{qp}(T)= \sum_i  d_i \int \frac{d^3p}{(2 \pi)^3} E_i f_i(E_i) + B(m_i(T)) 
  = \sum_i \epsilon_{kin}^{i}(m_i,T)+B(m_i(T)).
\end{eqnarray}
In the model there are therefore two unknown functions, $g(T)$ and $B(T)$, but they are not independent, they
are related through the thermodynamic consistency relationship (\ref{consistenza}). Therefore, only one
function needs to be determined, which we do by imposing the condition:
\begin{equation}  
  \epsilon_{qp}(T)=\epsilon_{lattice}(T)  .
\end{equation}
Following Ref. \cite{BHH98}, we derive both sides of the above equation with respect to $T$, then by using equation (\ref{gap}) and recalling that $m_i(T)=\alpha_i g(T)T$, we obtain:
\beq
\label{g_T}
  \frac{dg(T)}{dT}\bigg|_{\mu=0}=g\cdot 
  \left \{ 
  \frac{\frac{d\epsilon_{lattice}}{dT}-\frac{\partial \epsilon_{kin}}{\partial T}-\sum_i [ \frac{\partial \epsilon_{kin}^i}{\partial m_i}\frac{dm_i}{dT}+\frac{\partial B}{\partial m_i}\frac{dm_i}{dT}]}
       {\sum_i\frac{\partial \epsilon_{kin}^i}{\partial m_i}m_i+\sum_i\frac{\partial B}{\partial m_i}m_i}
  \right \},
\eeq
where $\epsilon_{kin}=\sum_i\epsilon_{kin}^i$. Therefore, by performing the fit to the lattice energy density, we can solve the differential equation for $g(T)$, given the initial value $g(T_0)=g_0$.
The function $g(T)$ obtained in this approach for $T>T_c$ is the same as the one obtained in other models, in which a parameterized form for $g(T)$ is assumed, and the parameters are fixed through a fit to the lattice data (see for example \cite{PKPS96}):
\beq
\label{Peshier_g_T}
g^2(T)=\frac{48 \pi^2}{(11 N_C-2 N_f)\ln{\left[\lambda(\frac{T}{T_C}-\frac{T_S}{T_C})\right]^2}}.
\eeq

In our model, $g(T)$ is obtained numerically from Eq.(\ref{g_T}). If we want to express it in terms of the above parameterization
(\ref{Peshier_g_T}), we obtain $\lambda=17.7$ and $T_S/T_C=1.15$ from a fit to the results of the 
Ref. \cite{hotQCD}\footnote{Notice that preliminary results for some observables have been published by the
hotQCD collaboration with the new hisq action \cite{Soldner:2010xk,Bazavov:2010pg}, which has a small taste
violation at low temperatures and small cutoff effects at large temperatures. However, for our analysis we use
the results obtained with the p4 action, since the energy density data are not yet available for the hisq
action.}, and $\lambda=2.6$ and $T_S/T_C=0.57$ from a fit to the results of the
Wuppertal-Budapest collaboration \cite{Borsanyi:2010cj}. 

We recall here that the values of $T_c$ are different in the two sets of lattice data, more specifically we
have used $T_c= 175$ MeV for the hotQCD case and $155$ MeV for the WB one.

\begin{figure}
\hspace{1.5cm}
\scalebox{.4}{\includegraphics{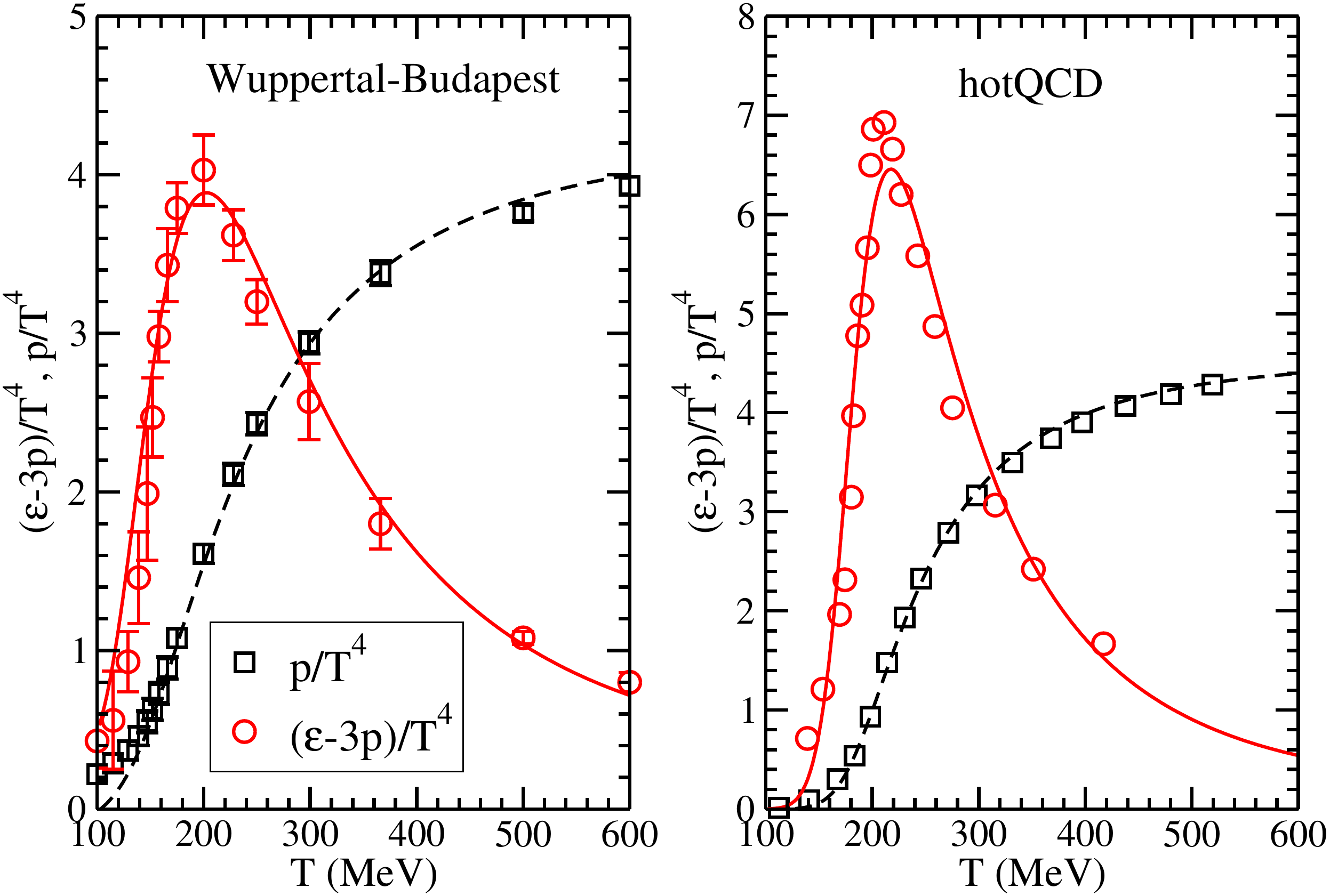}}
\caption{Lattice data for the pressure and trace anomaly as functions of the temperature, together with the quasi-particle model curves. The
results of the Wuppertal-Budapest collaboration are from
Ref. \cite{Borsanyi:2010cj},  the other lattice results
are taken from Ref. \cite{hotQCD}.}
\label{fig1}
\end{figure}

We perform our fit to the lattice data for the energy density; 
in Fig. \ref{fig1} we show the good agreement between our curves and lattice results for other quantities like
pressure and trace anomaly, due to thermodynamic consistency.
\begin{figure}
\hspace{-.5cm}
\scalebox{.55}{\includegraphics{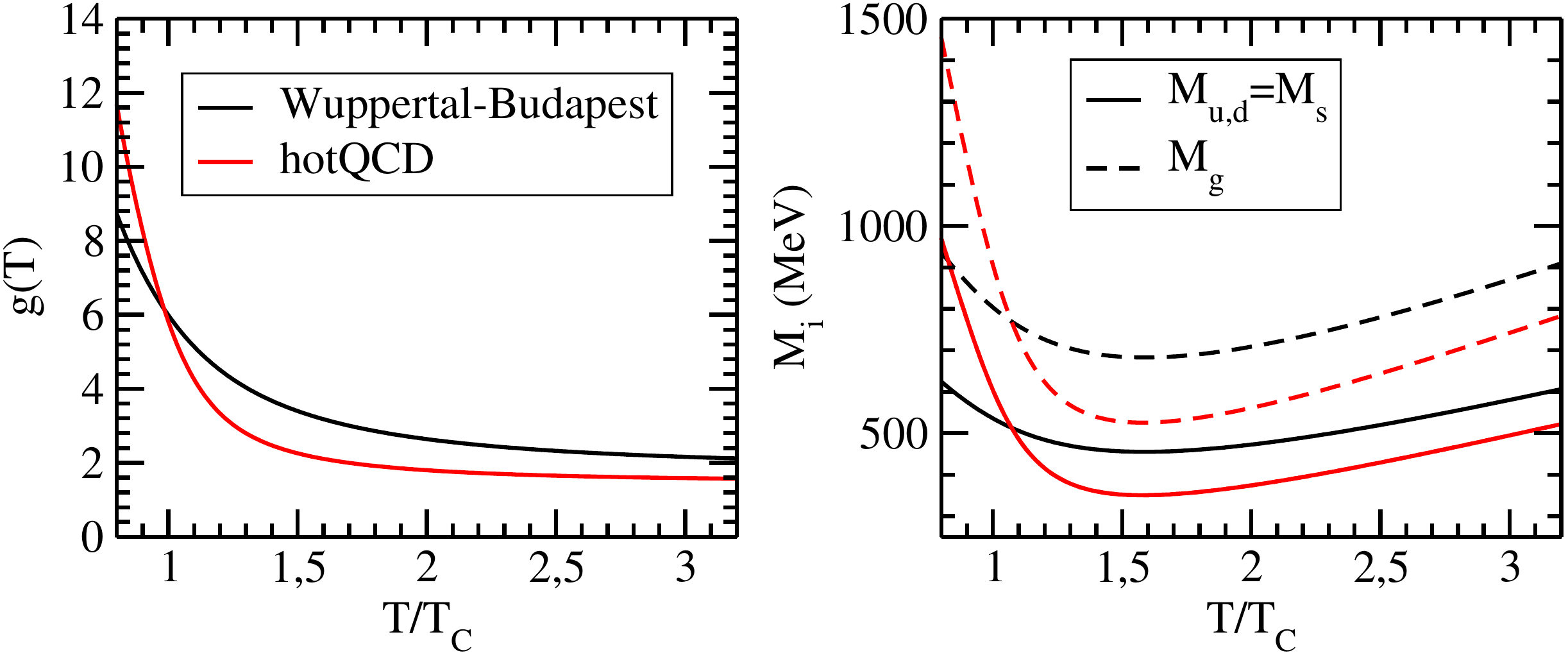}}
\caption{Left panel: temperature-dependent coupling $g(T)$ as a function of $T/T_C$.
Right panel: quark and gluon quasi-particle masses as functions of $T/T_C$. We used $m_{0s}=0$ in the fit: the effect of a finite current quark mass for the strange quark is negligible for the description of thermodynamic quantities. In both panels, the red curves
correspond 
to a fit of the lattice results from the hotQCD collaboration \cite{hotQCD}, the black ones to a fit of the
Wuppertal-Budapest collaboration results \cite{Borsanyi:2010cj}.}
\label{fig2}
\end{figure}

In Fig. \ref{fig2} we show the temperature dependence of the coupling $g$ and of the quark and gluon masses 
that we obtain from this procedure. As we can see, the slower increase of the energy density and
the smaller values at hight $T$ in the Wuppertal-Budapest (WB) data with respect to the hotQCD ones is reflected into larger quark and gluon masses. At small temperatures, as $T \rightarrow T_c$, the energy density from the Wuppertal-Budapest collaboration is larger than the hotQCD one, and as a consequence in the last case quark and gluon masses have a stronger $T$ dependence resulting in larger masses in this temperature regime.

We notice that, at sufficiently high temperature, $m \sim T$, as we can expect because $T$ remains
the only scale of the problem. When we are approaching the phase transition, there is instead a tendency
to increase the correlation length of the interaction: the quasiparticle model tells us that this can be
described as a plasma of particles with larger masses. 
This essentially determines the fall and rise behavior of $m(T)$ seen in all quasiparticle model
fits to lattice data in $SU(N)$ gauge theories including the $SU(3)$ case for QCD. 
In Fig. \ref{fig2} we can see a quite smoother behavior of the masses associated to the Wuppertal-Budapest
case, indicating that the strength of such correlation is significantly reduced when lattice simulations are performed at the physical quark masses and the continuum limit is taken.

\begin{figure}
\begin{center}
\scalebox{.4}{\includegraphics{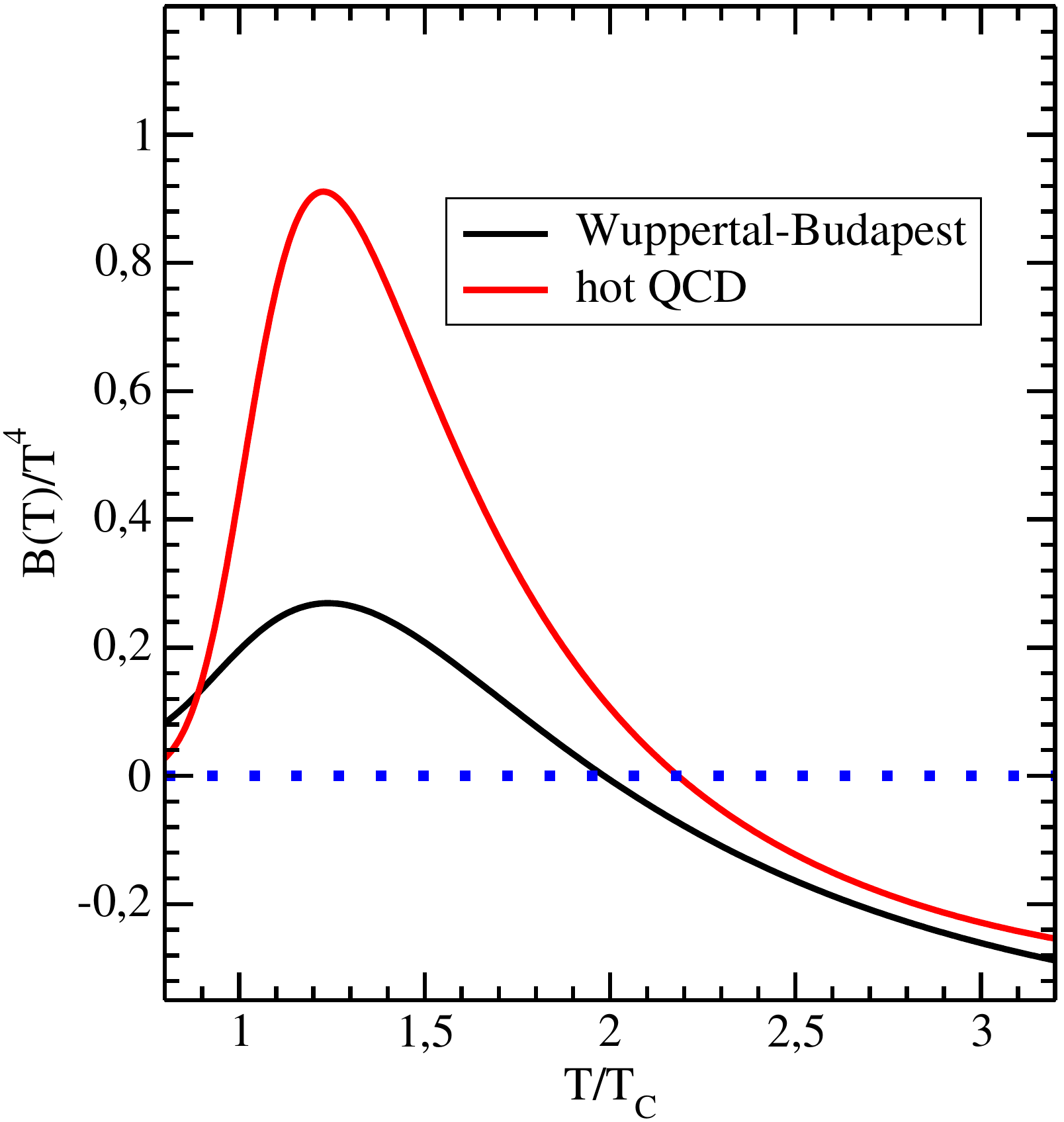}}
\caption{$B(T)$ as a function of $T/T_C$.}
\label{fig3}
\end{center}
\end{figure}

\begin{figure}
\begin{center}
\scalebox{.4}{\includegraphics{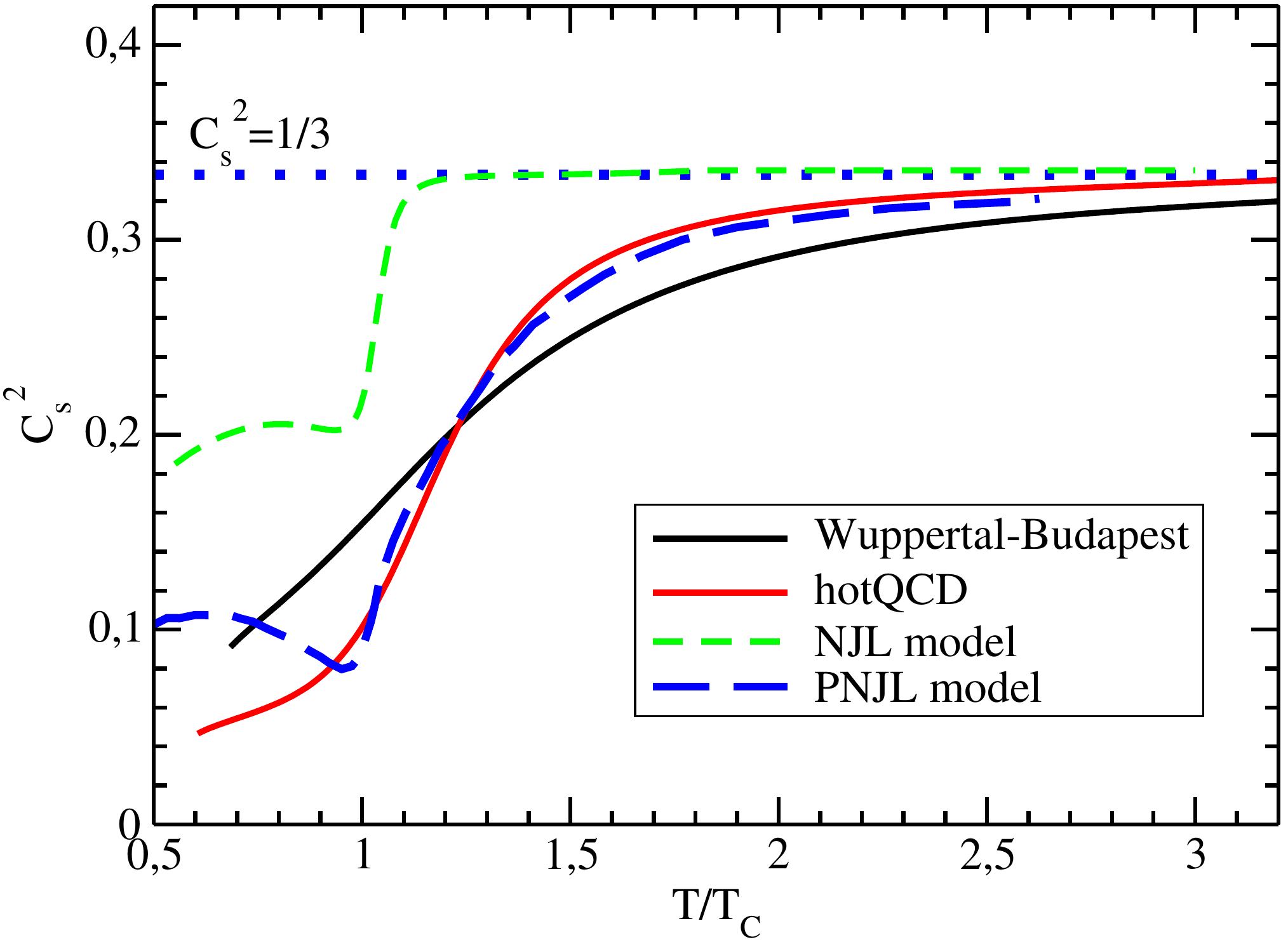}}
\caption{Speed of sound squared as a function of $T/T_C$. The red curve corresponds to the hotQCD data, the
black one to the Wuppertal-Budapest data, the short-dashed one to the NJL model and the long-dashed one to
the PNJL model fit to the hotQCD lattice data.}
\label{fig4}
\end{center}
\end{figure}

The function $B(T)$ can be obtained from the difference between the lattice energy density and the kinetic
contribution 
from our model: $B(T)=\epsilon_{lattice}(T)-\sum_i \epsilon_{kin}^i(T)$.  We show its temperature dependence in
Fig. \ref{fig3} for the two different fits to the lattice results. $B(T)$ provides another measure for
non-perturbative physics which cannot be absorbed into effective quark and gluon masses. As it is evident from
Fig. \ref{fig3},  in the vicinity of the phase transition the $B(T)$ corresponding to the hotQCD lattice data
(red curve) is much larger than the Wuppertal-Budapest one (black curve). This mainly reflects the discrepancy in the
interaction measure $(\epsilon-3P)/T^4$ between the results of the two collaborations (see Fig. 17 of Ref.
\cite{Borsanyi:2010cj}) .
>From the energy density and the pressure one can easily evaluate the speed of sound squared, defined as
\beq
C_{s}^{\,2}=\frac{dP}{d\epsilon};
\eeq
we show it for completeness in Fig. \ref{fig4}. The curve corresponding to the hotQCD fit displays a steeper rise in the vicinity of the transition temperature, and approaches the Stefan-Boltzmann limit
much faster with respect to the curve corresponding to the Wuppertal-Budapest fit.
For comparison, in Fig. \ref{fig4} we show the speed of sound as calculated for the Nambu-Jona-Lasinio (NJL) model,
the set of parameters used are those taken by Buballa \cite{BuballaRep,Plumari:2010}.
The speed of sound calculated in the NJL model has a rapid increase near the critical temperature and 
for $T \approx 1.2 T_C$ it converges towards the characteristic value of a relativistic massless gas.
This is due to the fact that in this model the trace anomaly is very small for $T > T_C$, see
\cite{Plumari:2010}.
However if the Polyakov loop dynamics is added the behavior of the sound velocity $c_s(T)$ is
very well reproduced as shown by the long dashed line taken from Ref. \cite{PNJL-cs}, which is fitted to the hotQCD
data. 

\section{Transport coefficients}

Once the properties of quark and gluon quasi-particles are fixed to reproduce the lattice data, we can use the
model to compute quantities that cannot be evaluated on the lattice, such as for example transport
coefficients. 
To study their thermal properties we use the method 
described by C. Sasaki and K. Redlich \cite{eta_quasi_part}, which is based 
on the relativistic kinetic theory formulated in the relaxation time approximation.
The formulas for the viscosities $\eta$ and $\zeta$ are derived for a quasi-particle 
description with bosonic and fermionic constituents (see \cite{eta_quasi_part} for details):
\beq
\eta=\frac{1 }{15T}\sum_i d_i \int \frac{d^3p}{(2 \pi)^3} \, \tau_i \, \frac{\vec{p}^{\,\,4}}{E^2_i} \, f_i \,
\left(1 \mp f_i \right) ;
\label{shear_eq}
\eeq

\begin{eqnarray} 
\label{bulk_eq}
\zeta=-\frac{1}{3T}\sum_i d_i 
\int \frac{d^3p}{(2 \pi)^3} \,\tau_i \; \frac{m_i^2}{E_i} \; f_i \;
\left(1 \mp f_i\right)
\left(\frac{\vec{p}^{\,\,2}}{3E_i}-C_s^2 
\left[ E_i - T\frac{\partial E_i}{\partial T} \right]
\right).
\end{eqnarray}
In the relaxation time approximation, both shear and bulk viscosities for the 
particles $i=u,d,s,g$ depend on their collisional relaxation time $\tau_i$.
The latter is given by the thermal average of the in-medium cross 
section $\sigma$ (describing total elastic scattering of medium constituents) 
times the relative velocity of two colliding 
particles $v_{rel}$ \cite{relax_time}:
\beq
\tau^{-1}_{i}(T)=n_{i}(T) \langle v_{rel} \: \sigma\rangle(T);
\eeq
$n_{i}$ is the density of the particle $i$.
The in-medium cross sections for quark-antiquark, 
quark-quark and antiquark-antiquark scattering processes were studied in detail in 
Ref. \cite{scat} within the NJL model for two different flavors, including $1/N_C$ 
next-to-leading order corrections. These results incorporate dominance of the 
scattering on large angles and take into account a possible occupation of particles in the final state, see Ref. \cite{Sasaki:2008um}.

The QCD calculations of the relaxation time $\tau$ of partons require the sum of infinitely many 
diagrams already in the lowest 
order in the running coupling constant $g$. 

A first evaluation of the hard thermal loops resummation \cite{rob,smilga} has suggested a parton width  
$\tau^{-1}$ going as $\sim g^2 T \ln(1/g )$. Thus in recent papers \cite{PC05,peshier2,KTV2010} within a quasiparticle approach the following parametrizations have been used for quarks and gluons: 
\begin{eqnarray}
\tau_q^{-1} &=& 2 \, \frac{N_C^2-1}{2 N_C} \, \frac{g^2 T}{8 \pi} \, \ln{\frac{2 k}{g^2}}, \nonumber \\
\tau_g^{-1} &=& 2 N_C \, \frac{g^2 T}{8 \pi} \, \ln{\frac{2 k}{g^2}},
\label{relax}
\end{eqnarray}
where $g$ is the coupling obtained by solving Eq. (\ref{g_T}) and $k$ is a parameter 
which is fixed by requiring that $\tau_i$ yields a minimum of one for the quantity $4\pi\eta/s$.

We point out that such expressions do not recover the leading-order pQCD limit where $\tau$ is proportional to $g^4\,ln(g^{-1})$ \cite{Arnold_1} and not $g^2\,ln(g^{-1})$. However, a pure $g^4$ 
parametric dependence is strictly valid in the weak coupling limit for $T>> \Lambda_{QCD}$, a condition that
is not fully satisfied in our temperature range. Nevertheless in Ref.\cite{Arnold_2} it has been shown that
parametric corrections to the leading order are subdominant and give at most a $g$ power correction at least when $m_D < T$.
One may still argue that in our temperature range, especially at $T<2T_c$ one is far from a weak coupling limit and the last condition is not fulfilled. Nevertheless we will consider the relaxation time expressions in the weak coupling limit showing that in the asymptotic limit our expression for $\eta/s$ can recover
the correct pQCD limit. Indeed, considering both the $g^2$ and $g^4$ dependences for $\tau^{-1}$ will be particularly interesting when more accurate lQCD calculations for the shear
viscosity will become available, allowing to infer the parametric dependence of $\tau$ within a quasiparticle
approach.
We have considered the following ansatz for the relaxation time as in Refs. \cite{Arnold_1, Arnold_2}
derived in a pQCD approach:

\begin{eqnarray}
 \tau_{g}^{-1} &=& C_{g}g^{4}T\ln{(a_{2}/g^{2})}  \nonumber \\
 \tau_{q}^{-1} &=& C_{q}g^{4}T\ln{(a_{2}/g^{2})}.
 \label{relax_AMY}
\end{eqnarray}

where $C_{q}/C_{g}=(N_{C}^{2}-1)/(2N_{C}^{2})=4/9$. 
The validity region of pQCD is certainly not clear, however we will show in the following that
the  $g^4$ pQCD parametric dependence of the relaxation time gives a significantly
different behavior as a function of the temperature.
We use these $\tau_{j}$ in order to recover, 
at large $T$, the dependence of the shear viscosity on temperature and coupling known from pQCD 
\cite{Arnold_1,Arnold_2}.

\begin{figure}
\begin{center}
\scalebox{.55}{\includegraphics{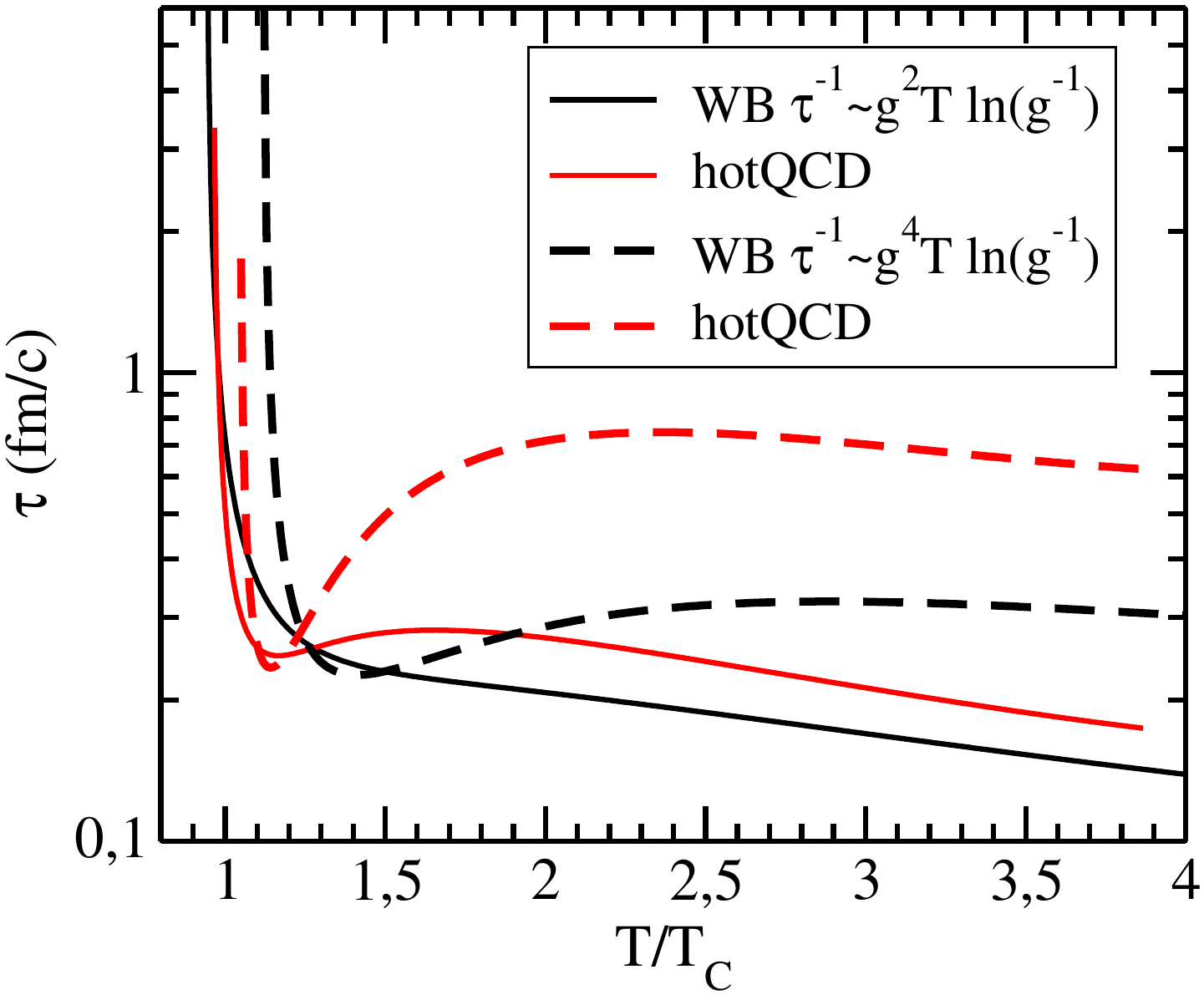}}
\caption{Relaxation time $\tau_g$ for gluons as a function of $T/T_C$. The red curves correspond to the hotQCD data, the black curves to the 
Wuppertal-Budapest ones. The solid lines correspond to Eq. (\ref{relax}), the dashed lines to Eq. (\ref{relax_AMY}).}
\label{fig_tau}
\end{center}
\end{figure}

In Fig. \ref{fig_tau} we show the relaxation time for gluons as a function of $T/T_C$
for the two different lattice fits and the two different relaxation times in 
Eq.(\ref{relax}) and Eq.(\ref{relax_AMY}) (discussed later in more detail).
The results shown in the figure are obtained using 
the same value of the parameter, $k=22$, similarly to \cite{KTV2010_gluon}.
As we can see for the $g^2$ parametric dependence (dashed lines) the relaxation time is a monotonically decreasing function of $T$ for both lattice fits, apart from a shallow minimum slightly above $T_c$ in the hotQCD case. The trend changes significantly for the $g^4$ dependence, that in both cases
generates a minimum in the relaxation time: this can be considered as an indication of a strong
increase of the interaction just above the phase transition temperature. We notice that the smoother phase transition dynamics of the WB lattice QCD data is reflected also in the relaxation time, by generating a
less deep minimum with respect to the hotQCD case.

\begin{figure}
\begin{center}
\scalebox{.5}{\includegraphics{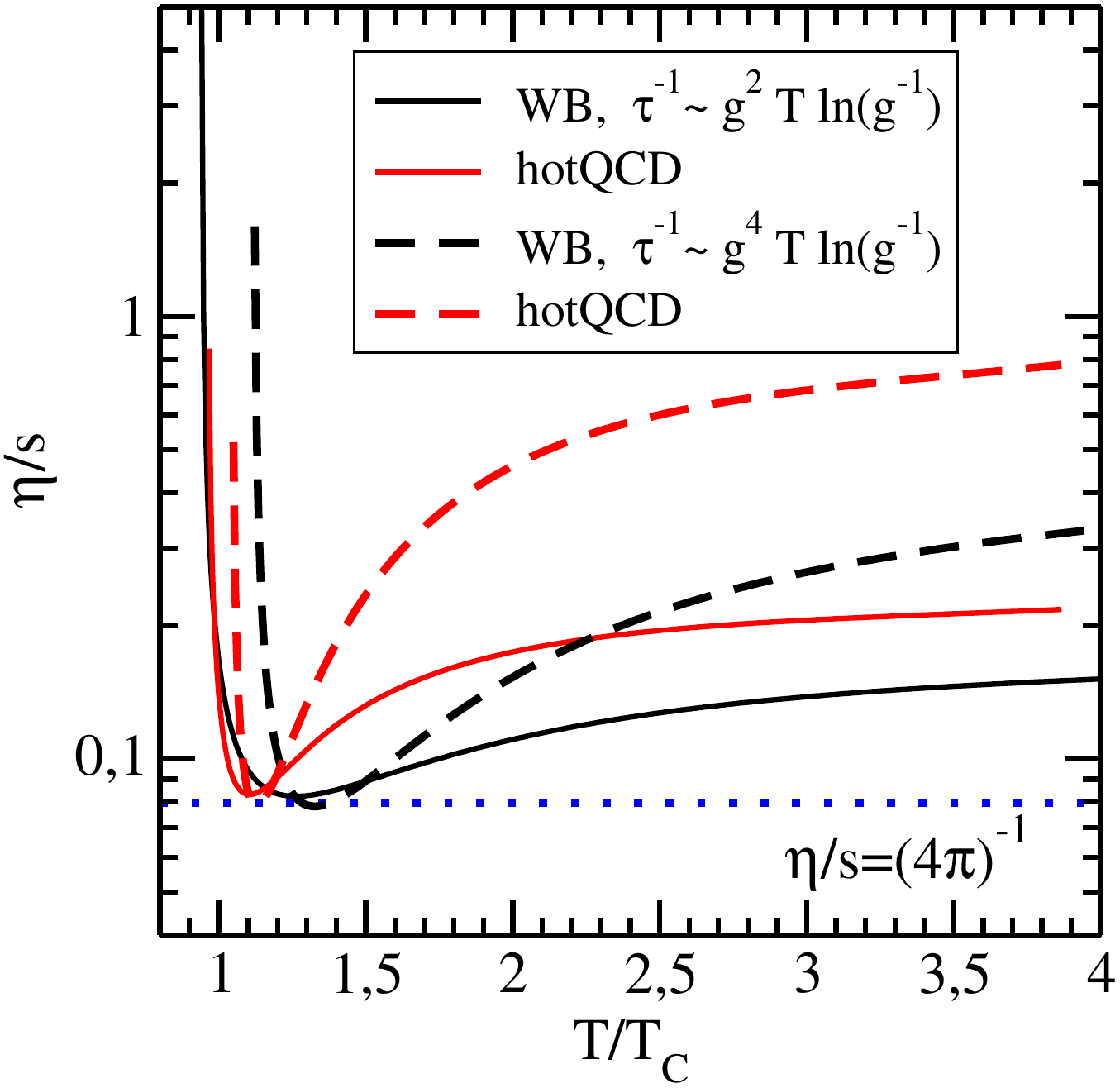}}
\caption{Shear viscosity to entropy density ratio $\eta/s$ as a function of $T/T_C$.
 The red curves correspond to the hotQCD data, the black curves to the Wuppertal-Budapest ones. 
 The solid lines correspond to a $g^2$ dependence of $\tau$, the dashed lines to a $g^4$ dependence of $\tau$. The dotted line is the $\eta/s=1/(4\pi)$ bound.}
\label{fig_shear}
\end{center}
\end{figure}

In Fig. \ref{fig_shear} we show the shear viscosity to entropy density ratio, $\eta/s$, as a function of $T/T_C$ for the $g^2$ case (solid lines) and the $g^4$ case (dashed lines). It is interesting to notice that, once $k$ is fixed to have a minimum $\eta/s=1/4\pi$, this minimum appears very close to $T_c$ and is followed by a divergency as  $T \to T_c$. 
This behavior is shared by all the cases considered. Of course the exact location
of the minimum goes beyond the parton quasiparticle approach, that breaks its vality for $T\leq T_c$,
however on general arguments a minimum is expected very close to $T_c$ \cite{Csernai:2006zz}, furthermore one also expects a large increase of $\eta/s$ below $T_c$, entering the hadronic phase.

At high temperature the viscosity corresponding to the Wuppertal-Budapest fit shows a 
slower increase with temperature with respect to the one corresponding to the hotQCD fit. In the case of the $g^2$ dependence of $\tau$, the increase in $\eta/s$ is quite mild and it seems that even at 
temperatures $T\sim 4 T_c$ this transport coefficient is much smaller than its estimate in pQCD, 
$\eta/s\simeq1$. This is in apparent contradiction with the behavior of the 
trace anomaly, which decreases quite
rapidly with increasing $T$, approaching the perturbative estimate relatively early. 
Such a behavior can be mostly related to the uncertainty in the evaluation of the relaxation time $\tau$
and in particular its temperature dependence.
In fact, as mentioned before, 
Eq.(\ref{relax}) does not give the proper pQCD limit at very large temperature and may be questionable
already at moderate temperature $T \sim 4 T_c$ as those considered here. By using Eq.(\ref{relax_AMY})
not only the pQCD relaxation time is recovered, but also the correct limit
for the shear viscosity is guaranteed by Eq.(\ref{shear_eq}), as we discuss in the following.
In \cite{Arnold_2} the shear viscosity was obtained for $N_{f}=3$ 
and $N_{C} = 3$ at next-to-leading log order of the small running coupling $g$.

\begin{equation}
  \eta = \frac{\eta_{1} T^{3}}{g^{4}\ln{\frac{\mu_{*}}{gT}}}
  \label{eta_NLL}
\end{equation}
with $\eta_{1}=106.66$ and $\mu_{*}/T=2.957$. In our quasiparticle approach, from Eq.(\ref{shear_eq}) 
for the shear viscosity and employing
the previous ansatz for the relaxation time, we get

\begin{equation}
  \eta = \frac{\left[ \frac{G(T)}{C_{g}}+\frac{Q(T)}{C_{q}} \right] T^{3}}{2 g^{4}
\ln{\frac{\sqrt{a_{2}}}{gT}}}
  \label{eta_qpmodel}
\end{equation}

where $G(T)$ and $Q(T)$ are two dimensionless functions of the temperature rispectively 
for the gluon and the quark contribution. 
Due to the fact that the coupling constant $g(T)$ in this model is a logarithmic decreasing function 
with the temperature we have that the above functions approach their asymptotic limits

\begin{eqnarray}
 G(T\to \infty) &=& d_{g}\,4\,\Gamma(4)\zeta(4)/(30\pi^{2})\approx 1.404 \nonumber \\
 Q(T\to \infty) &=& d_{q}\,(7/2)\,\Gamma(4)\zeta(4)/(30\pi^{2})\approx 2.763 \nonumber
\end{eqnarray}

Therefore the analytic behavior of $\eta/s$ in Eq. (\ref{eta_qpmodel}) shows that the quasi-particle approach
in the relaxation time approximation is able to reproduce asymptotically the correct pQCD limit.
At very high temperatures, comparing Eqs. (\ref{eta_NLL}) and (\ref{eta_qpmodel}), we obtain the 
following values for the coefficients of the relaxation time $C_{g}=3.573 \cdot 10^{-2}$, 
$C_{q}=(4/9)C_{g}=1.588 \cdot 10^{-2}$ and $a_{2}=(\mu_{*}/T)^{2}$. In the $T \rightarrow \infty$
limit, with these coefficients we obtain the same perturbative limit of Eq.(\ref{eta_NLL}) for both the hotQCD 
and Wuppertal-Budapest fits, even if the quickness with which this limit is reached
can be quite different. 
Notice that in the asymptotic limit $g(T)T \rightarrow \infty$, the same pQCD limit
is recovered independently of the $a_2$ values, because in such a limit 
$ln(\sqrt{a_2}/gT) \rightarrow -ln\, gT$.

Choosing $\mu_*=\pi T$ as in Ref. \cite{Arnold_2}, the shear viscosity shows a minimum above $T_{c}$ for both
fits and the value of $\eta/s$ remains close to the perturbative limit in fact we get a minimum
$(\eta/s)_{min}\approx 0.68$ for the hotQCD fit and $(\eta/s)_{min}\approx 0.66$ for
the Wuppertal-Budapest fit. 
This result however is in strong disagreement with recent lattice QCD calculations that show that the ratio $\eta/s \approx 1/(4 \pi)$ at temperatures
close to the critical temperature, as well as with all the estimates
of $\eta/s$ derived from heavy-ion collisions at RHIC energies within hydrodynamics 
\cite{Romatschke:2007mq, Song:2008si} and/or the parton transport model
\cite{greco_cascade, Greiner_cascade}.

Therefore, in a way similar to the $g^2$ parametrization, the value of the minimum in $\eta/s$ depends on the value of the $a_{2}$ coefficient in the logarithm. If we choose $a_{2}=25$ (close to the value used for the other ansatz for $\tau$) while keeping the same
values of the coefficients $C_{g}$ and $C_{q}$ in order to
have the correct pQCD limit, we obtain that the minimum of $\eta/s\sim1/(4 \pi)$, in agreement with the current estimates in the $1-2\, T_c$ range from heavy-ion collisions
phenomenology and first lattice QCD analyses.
We see that the $g^4$ ansatz produces a significantly stronger and more rapid rise
of $\eta/s$ (dashed lines in Fig.\ref{fig_shear}). There is however still a noticeable
difference between $\eta/s$ in the WB case, that is always smaller than
that calculated for the hotQCD fit. This is in agreement with the general slower convergence of the
thermodynamics towards the perturbative gas limit in the WB case, that is reflected by  the different behavior of the coupling constants, see left panel of Fig.\ref{fig2}.
As shown in Fig.\ref{fig_shear}, we obtain that the position of the minimum for the dashed curves is very close to the one obtained in
the previous parametrization for the relaxation time Eq.(\ref{relax}), and for increasing temperatures 
we have that $\eta/s$ is a monotonically increasing function for both fits. It asymptotically 
approaches the pQCD limit of Eq.(\ref{eta_NLL}), while for the parametrization of Eq.(\ref{relax}) 
we have that $\eta/s$ never reaches this limit.
The hotQCD case gives a $4\pi \eta/s \sim 10$ (relatively close to the perturbative limit) already at $T \sim
4 T_c$ hence predicting a nearly perturbative plasma in the initial stage created at the LHC.
The WB case instead shifts this limit at much higher temperatures and at $T\sim 4\, T_c$ we still have a 
$4\pi \eta/s \sim 4$. The latter could be the reason behind a similar amount of observed elliptic flow
at RHIC and LHC energies \cite{Collaboration:2011vk}

\begin{figure}
\begin{center}
\scalebox{.5}{\includegraphics{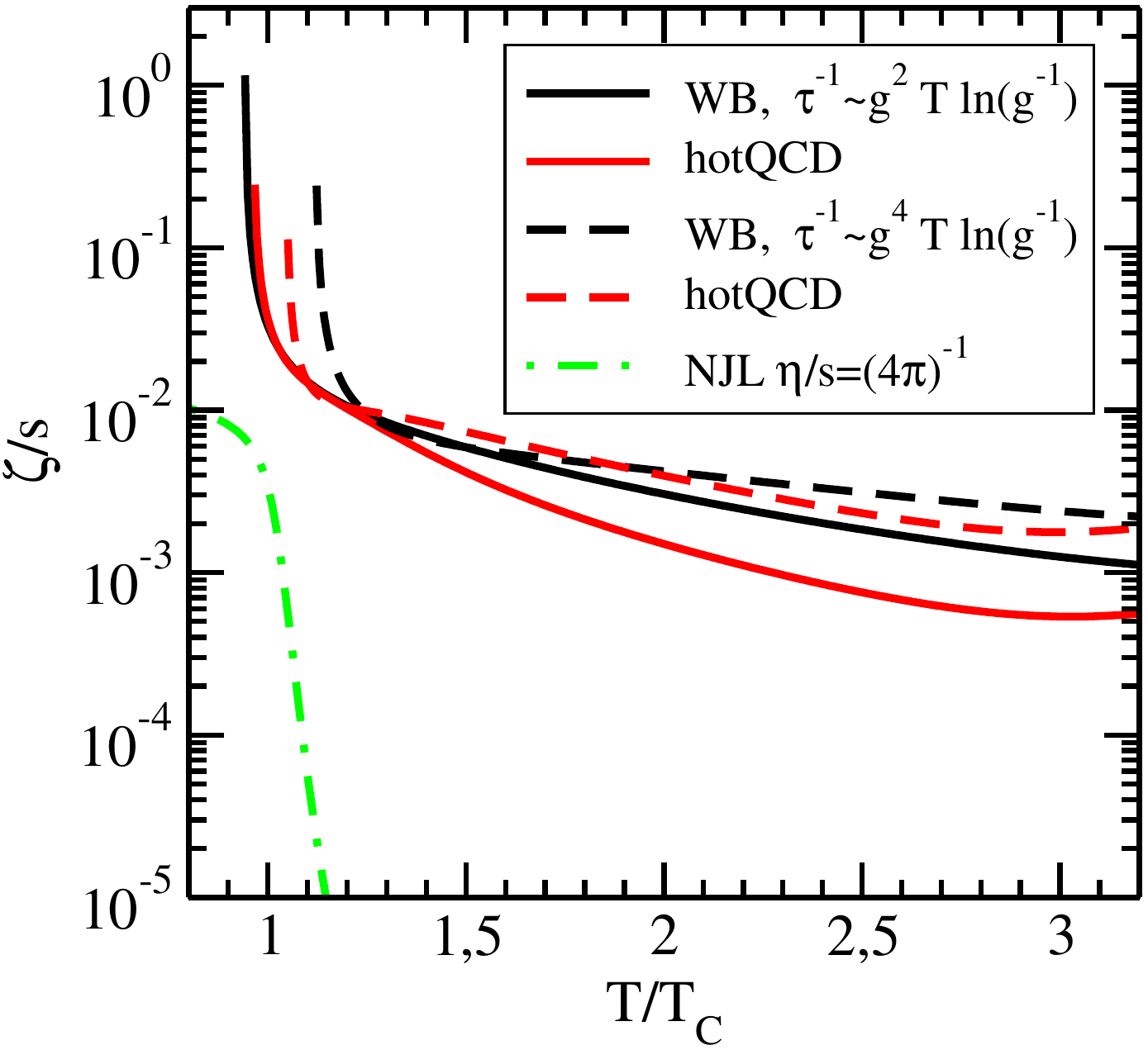}}
\caption{Bulk viscosity to entropy density ratio $\zeta/s$ as a function of $T/T_C$.
The solid lines are for the $\tau$ of Eq.(\ref{relax}), the dashed lines are for the $\tau$ of Eq.(\ref{relax_AMY}). The red curves correspond to the hotQCD data, the black curves to the Wuppertal-Budapest ones. The dot-dashed curve corresponds to the NJL model with $\eta/s=(4\pi)^{-1}$, see the text for details.}
\label{fig_bulk}
\end{center}
\end{figure}

To evaluate the bulk viscosity, Eq.(\ref{bulk_eq}), we have used the crude approximation of small relaxation time as for the shear viscosity. However, the bulk viscosity has a different origin
and the processes involved could lead to a different $\tau$ as discussed in Ref.\cite{Arnold_bulk}.

In Fig. \ref{fig_bulk} we show the bulk viscosity to entropy density ratio $\zeta/s$ as a function of 
$T/T_C$. 
As we can see from Eq. (\ref{bulk_eq}), the bulk viscosity is more sensitive to the equation of state 
than the shear viscosity and the term $\left[ E_i - T\frac{\partial E_i}{\partial T} \right]$ 
goes to $\vec{p}^{\,\,2}/E$ for $T \gg T_C$; therefore, at high temperatures the behavior of the bulk viscosity depends on how $C_s^2 \to 1/3$.
Again the difference can be traced back to the different quasi-particle masses, but in general the differences
among the cases considered are much smaller than in the case of the shear viscosity.
In general, there is a compensation between the behaviors of $g(T)$ and $C_s(T)$ with a tendency
to have a smaller $\zeta/s$ for the hotQCD case, that can be traced to the faster increase
of $C_s$ towards $1/3$.

For comparison, in Fig. \ref{fig_bulk} we show the ratio $\zeta/s$ for the NJL model with the same set of parameters used for the speed of sound, see Fig. \ref{fig4}. In this calculation the relaxation time is chosen 
in such a way to keep the ratio $\eta/s$ constant and fixed to the lower bound $\eta/s=(4\pi)^{-1}$.
As we can see, the bulk viscosity for the NJL model is much smaller by orders of magnitude than the one calculated using the quasi-particle 
model and it has a rapid decrease for $T>T_C$. 
This is a direct consequence of the fact that a large mass ($m \sim 300-400$ MeV) in NJL is related 
only to the chiral phase transition which modifies the speed of sound only for $T \leq 1.2 T_C$ 
and approaches the relativistic limit of $C_s^2=1/3$ just
above $T_c$, as shown in Fig \ref{fig4}. Note that such a large difference does not depend on the particular relaxation time used, because $\zeta/s$ for $T > 1.2 T_C$ is already several order of magnitudes smaller with
resepect to the quasi-particle model.

\section{Susceptibilities}
\label{qsusc}
We conclude our analysis by showing the predictions on quark number susceptibilities that we obtain in the quasi-particle model for the two different fits to the lattice data.
Quark number susceptibilities are a very useful tool to understand the nature of the degrees of freedom in the vicinity of the QCD phase transition. They have been studied both on the lattice
\cite{hotQCD2,Borsanyi:2010bp} and using phenomenological models \cite{susce0,susce}
The ones that we address here are defined in the following way:

\begin{displaymath}
{c_{2}^{uu}=\left.\frac{T}{V}\frac{\partial^2\ln Z}{\partial\mu_{u}^{2}}\right|_{\mu_i=0}~~~~~~
\chi_{2}^{s}=\left.\frac{T}{V}\frac{\partial^2\ln Z}{\partial\mu^{2}_{s}}\right|_{\mu_i=0}}.
\end{displaymath}

\begin{figure}
\begin{center}
\scalebox{.4}{\includegraphics{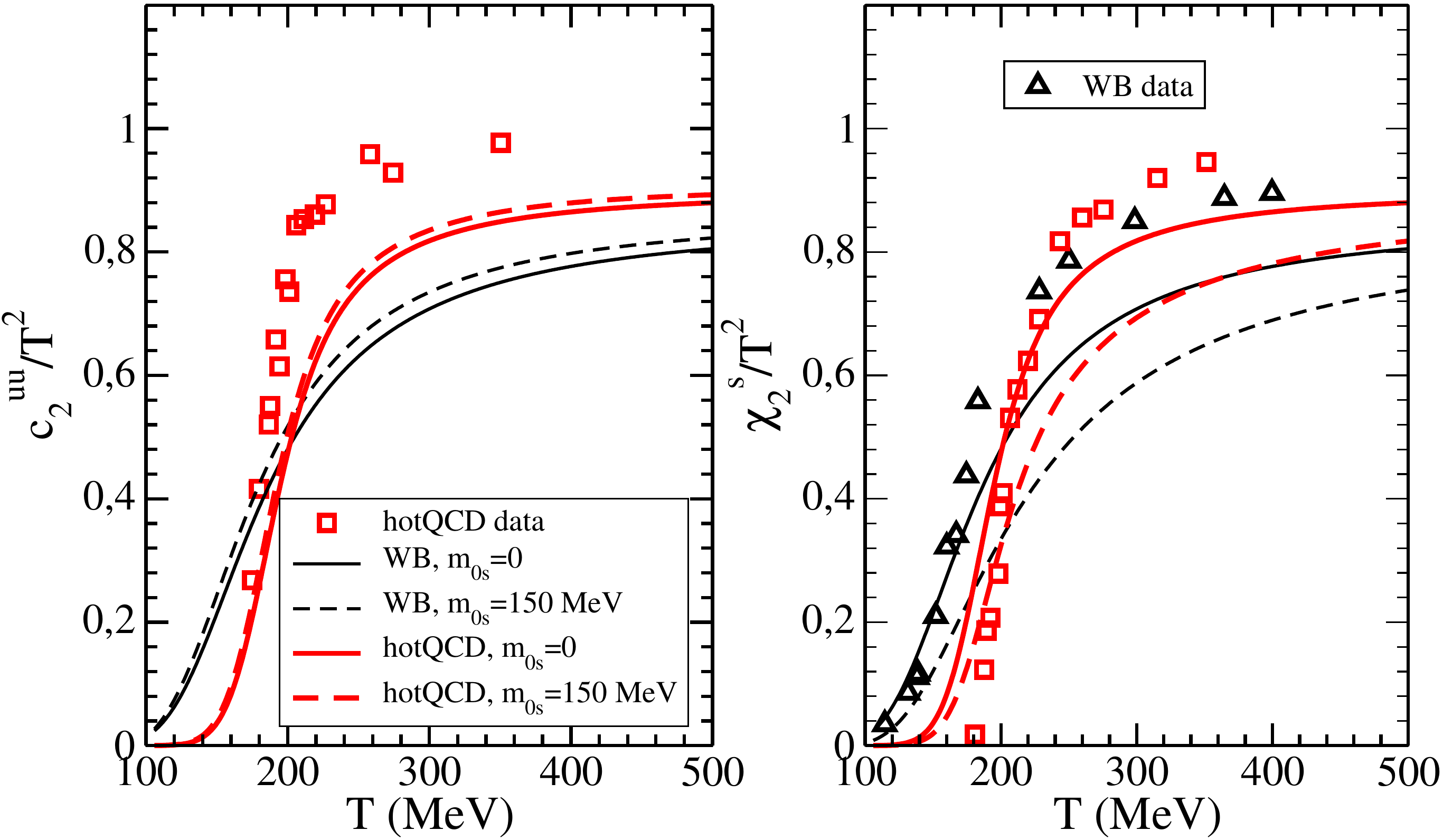}}
\caption{Light (left panel) and strange (right panel) quark number susceptibilities as functions
of the temperature. In both panels, the green lines correspond to the hotQCD fit, the black lines
to the Wuppertal-Budapest one. The solid curves have been obtained by taking equal masses for the light and strange quarks, the dashed curves correspond to a heavier strange quark mass, with a shift $m_{0s}=150$ MeV. The hotQCD data are taken from Ref. \cite{hotQCD2}, the WB ones from Ref. \cite{Borsanyi:2010bp}.}
\label{fig_susc}
\end{center}
\end{figure}

The results published so far by the hotQCD collaboration on light quark susceptibilities \cite{hotQCD2} seem to indicate that quarks behave as almost massless, non-interacting quasiparticles immediately above $T_C$.
This can be seen in Fig. \ref{fig_susc}, where we plot the predictions of the quasi-particle model
for light and strange quark number susceptibilities, in comparison to the available lattice data
from the hotQCD and the Wuppertal-Budapest collaborations (light quark number susceptibilities are not yet available in the latter case). 

Earlier works have calculated the quark susceptibilities using a quasi-particle model. In particular
in Ref. \cite{susce0} the $g(T)$ of the quasiparticle model has been directly fitted to $c_2^{uu}(T)$, then
predicting $c_4(T)$. Such a study has shown that, in principle, it is possible to account for
the quark susceptibilities in a quasi-particle approach. 
Our study is however meant to check deeply the consistency of the present quasi-particle model
not fitting the susceptibilities but calculating them from the fit to the energy density (see Fig. \ref{fig2}).
We can see from Fig. \ref{fig_susc} that, for the hotQCD fit (dashed line) we underestimate the data. Indeed, the 
lattice data already reach the ideal gas limit for temperatures slightly above $T_C$, thus leaving little space for 
a thermal quark mass. In fact, such observable is very sensitive to the quark masses that we use as inputs. This apparent inconsistency between the description of QCD thermodynamics and susceptibilities within the quasi-particle model has been observed before \cite{susce0}

Apart from the comparison to the hotQCD data, our results show that 
we should expect for the WB case a smoother behavior of $c_{2}^{uu}(T)$ across $T_c$; in fact
 we can see in Fig. \ref{fig_susc} (left) that the susceptibilities corresponding to the 
Wuppertal-Budapest fit (solid line) are significantly lower than the hotQCD ones even at large 
temperatures, due to a difference of $\sim 100$ MeV in the corresponding quark masses (see Fig. \ref{fig2}). 
Notice that the hotQCD lattice data shown in Fig. \ref{fig_susc} are obtained for heavier-than-physical 
quark masses and at finite lattice spacing: they might change once the continuum limit is taken and 
physical quark masses are used in the simulations. 

In Fig. \ref{fig_susc}  we also show the results taking into account a finite strange quark mass
($m_{0s}= 150$ MeV). This in principle makes the model more realistic. We have performed the same
procedure of fit of the energy density as described in Section 2 but for the non vanishing $m_{0s}$.
For the pressure, interaction measure, gluon and ligth quark masses we get essentially the
same result as in the vanishing $m_{0s}$ case. We can see this also for $c_{2}^{uu}(T)$ in Fig. \ref{fig_susc} (left), 
where the $m_{0s}= 150$ MeV case is shown by dashed lines and is quite similar to the solid lines corresponding 
to $m_{0s}=0$.
Nevertheless, the effect is significant for the strange quark susceptibility $\chi_{2}^{s}$ as one would
expect, but this increases the difference with respect to the available lattice QCD results
for both WB and hotQCD. This is just another confirmation that the source of the discrepancy
is the large value of the quark thermal masses. Therefore, if the trend of the present lattice data should be 
confirmed, in order for a quasi-particle description to hold, the thermal mass of quarks should be much 
lighter than the one we obtain from our fit. This would imply that the perturbative relationship Eq. (\ref{mqg}) 
between the quark and gluon thermal masses should not hold anymore, or some other dynamical mechanism
should be included in the quasiparticle description. One candidate could be the Polyakov
loop, whose dynamics can probably account for the gluon dynamics close to the phase transition.

\begin{figure}
\begin{center}
\scalebox{.4}{\includegraphics{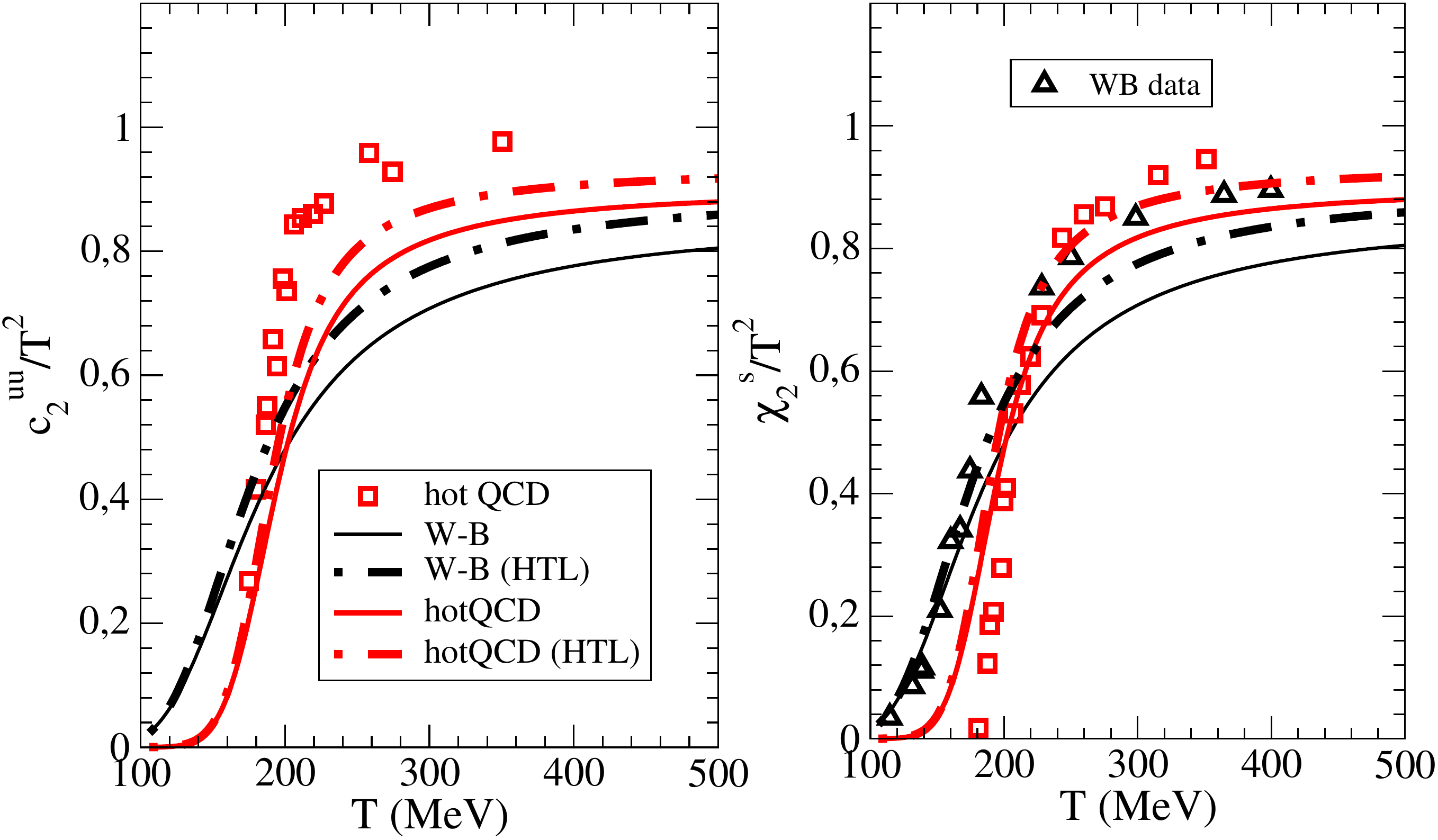}}
\caption{Light (left panel) and strange (right panel) quark number susceptibilities as functions
of the temperature. In both panels, the red lines correspond to the hotQCD fit, the black lines
to the Wuppertal-Budapest one. The solid curves are the ones presented in Fig. \ref{fig_susc}. 
The dot-dashed curves are obtained by using HTL masses for quarks and gluons. The hotQCD data are taken from Ref. \cite{hotQCD2}, the WB ones from Ref. \cite{Borsanyi:2010bp}.}
\label{suscHTL}
\end{center}
\end{figure}

Another possibility could be to use the quark and gluon masses that come from the Hard Thermal Loop formalism. 
In this case, quark masses would be smaller than the ones we obtain from our fit, while gluons would be heavier 
as given by the following relations:
\begin{eqnarray}
& &  m_g^2 = \frac{1}{3} g^2
    \left( N_c+ \frac12\, n_{\!f} \right) T^2 \, ,
  \nonumber \\
& &  m_{u,d,s}^2 = \frac{1}{2} \frac{N_c^2-1}{8N_c} g^2 \, T^2 \, 
\end{eqnarray}
In Fig. \ref{suscHTL} we compare the susceptibilities that we obtain
with the HTL relations (dash-dotted lines) to the pQCD ones from Fig. \ref{fig_susc} (solid lines). In both cases, a fit to the lattice QCD energy density has been performed. It is clear that the results obtained by using
the HTL ratio between quark and gluon thermal masses are significantly closer to the lQCD data available for 
the hotQCD case, in particular for the strange susceptibility $\chi_2^s$.
\section{Conclusions}
We have given an interpretation of recent lattice data for QCD thermodynamics in terms of quark and gluon 
quasi-particles with a thermal mass which we obtain from a fit to the lattice results. A thorough comparison 
between the results obtained from the fit to the hotQCD and Wuppertal-Budapest data is performed. 
We find that the differences between the results of the two collaborations are reflected in the differences 
between the quasi-particle characteristics, such as masses and bag constant.
In particular, the fit to the Wuppertal-Budapest data implies larger masses but with a weaker temperature
dependence, which is associated to a weaker evolution of the running coupling.
This is reflected on the dynamical properties of the quasi-particle medium. 
The shear viscosity over entropy ratio corresponding to the Wuppertal-Budapest fit shows a 
slower increase with increasing temperature with respect to the one corresponding to the hotQCD fit, 
thus favoring a hydrodynamical approach also in the LHC temperature regime.
However we find that, once the quasi-particle properties are fixed to reproduce the energy density
and/or the interaction measure, the predictions for quark number susceptibilities 
underestimate the presently available lattice results for both the WB and hotQCD cases.
This seems to indicate that the thermal quark masses that we obtain from the fits to QCD thermodynamics are 
too large. This last result casts some doubts on the capability of the present quasi-particle model to correctly
describe the inner dynamical structure of lattice QCD and hints at the necessity to 
further augment the dynamical structure of a quasiparticle approach in order to have a safe and reliable
tool to understand the physics behind the lattice QCD data.
A HTL approach suggests a smaller quark to gluon mass ratio $m_q/m_g$, which reduces the discrepancy with 
lQCD data for the susceptibilities while mantaining the same level of agreement for the pressure 
and energy density. We notice that transport equation self-consistency associated to the present 
quasi-particle model can be easily derived in a way similar to Refs. \cite{BHH98,Plumari:2010}.
This will allow to go beyond the cascade approach that takes into account only collisions among massless 
partons \cite{greco_cascade,Greiner_cascade,Molnar_cascade}, including a field dynamics that can correctly reproduce 
the thermodynamics of lQCD data.
\section*{Acknowledgements}
The authors gratefully acknowledge fruitful discussions with R. Bellwied, M. Bluhm and K. Redlich.
This work was supported in part by funds provided by the Italian Ministry of Education, Universities and Reserach under the Firb Research Grant RBFR0814TT.

\end{document}